\documentclass[reprint, amsmath, amssymb, floatfix, aps, superscriptaddress, nofootinbib, nobibnotes,twocolumn]{revtex4-1}

\usepackage[utf8]{inputenc}
\usepackage{graphicx}
\usepackage{geometry}
\usepackage{natbib}
\usepackage{dblfnote}
\DFNalwaysdouble
\usepackage{slashed}
\usepackage{graphicx}
\usepackage{dcolumn}

\usepackage{color}
\usepackage{hyperref}
\usepackage{array}
\usepackage{acro}
\usepackage{mathtools}
\DeclareAcronym{PBH}{
  short = PBH ,
  long = primordial black hole ,
  short-plural = s,
  foreign-plural = {}
}
\DeclareAcronym{MW}{
    short = MW,
    long = Milky Way,
    foreign-plural = {}
}
\DeclareAcronym{DM}{
    short = DM,
    long = dark matter,
    foreign-plural = {}
}
\DeclareAcronym{IBD}{
    short = IBD,
    long = inverse beta decay,
    foreign-plural = {}
}
\DeclareAcronym{CC}{
    short = CC,
    long = charged-current,
    foreign-plural = {}
}
\DeclareAcronym{NC}{
    short = NC,
    long = neutral-current 
}
\DeclareAcronym{SNR}{
    short = SNR,
    long = signal-to-noise ratio,
    foreign-plural = {}
}
\DeclareAcronym{DSNB}{
    short = DSNB,
    long = diffuse supernova neutrino background
}
\DeclareAcronym{Super-K}{
    short = Super-K,
    long = Super-Kamiokande neutrino observatory
}
\DeclareAcronym{BBH}{
  short = BBH ,
  long = binary black hole ,
  short-plural = s,
  foreign-plural = {}
}
\DeclareAcronym{GW}{
  short = GW ,
  long = gravitational wave ,
  short-plural = s, 
  foreign-plural = {}
}
\DeclareAcronym{LIGO}{
  short = LIGO ,
  long = Advanced Laser Interferometer Gravitational-Wave Observatory ,
  short-plural = ,
}
\DeclareAcronym{FERMI-LAT}{
    short = FERMI-LAT,
    long = Fermi Large Area Telescope
}
\DeclareAcronym{INTEGRAL}{
    short = INTEGRAL,
    long = International Gamma-Ray Astrophysics Laboratory
}
\DeclareAcronym{JUNO}{
    short = JUNO ,
    long  = Jiangmen Underground Neutrino Observatory
}
\DeclareAcronym{NFW}{
    short = NFW,
    long = Navarro-Frenk-White
}
\DeclareAcronym{eES}{
    short = eES,
    long = electron elastic scattering
}
\DeclareAcronym{pES}{
    short = pES,
    long = proton elastic scattering
}
\DeclareAcronym{PSD}{
    short = PSD,
    long = pulse-shape discrimination
}
\DeclareAcronym{LS}{
    short = LS,
    long = liquid scintillator
}
\DeclareAcronym{PMT}{
    short = PMT,
    long =  photomultiplier tube,
    foreign-plural = {}
}
\DeclareAcronym{ISO}{
    short = ISO,
    long = cored isothermal
}
\DeclareAcronym{EIN}{
    short = EIN,
    long = Einasto
}
\DeclareAcronym{MNS}{
    short = MNS,
    long = Maki-Nakagawa-Sakata
}
\DeclareAcronym{EG}{
    short = EG,
    long = extragalactic $\gamma$-rays
}

\newcommand{\citeeq}[1]{Eq.~(\ref{#1})}
\newcommand{\citefig}[1]{Fig.~\ref{#1}}

\begin{document}

\title{Constraining primordial black holes as dark matter at JUNO}

\author{Sai Wang}
\email{wangsai@ihep.ac.cn}
\affiliation{Theoretical Physics Division, Institute of High Energy Physics, Chinese Academy of Sciences, Beijing 100049, China}
\affiliation{School of Physical Sciences, University of Chinese Academy of Sciences, Beijing 100049, China}

\author{Dong-Mei Xia}
\email{xiadm@cqu.edu.cn}
\affiliation{Key Laboratory of Low-grade Energy Utilization Technologies \& Systems of Ministry of Education of China, College of Power Engineering, Chongqing University, Chongqing 400044, China}

\author{Xukun Zhang}
\email{zhangxukun@ihep.ac.cn}
\affiliation{Theoretical Physics Division, Institute of High Energy Physics, Chinese Academy of Sciences, Beijing 100049, China}
\affiliation{School of Physical Sciences, University of Chinese Academy of Sciences, Beijing 100049, China}

\author{Shun Zhou}
\email{zhoush@ihep.ac.cn}
\affiliation{Theoretical Physics Division, Institute of High Energy Physics, Chinese Academy of Sciences, Beijing 100049, China}
\affiliation{School of Physical Sciences, University of Chinese Academy of Sciences, Beijing 100049, China}

\author{Zhe Chang}
\email{changz@ihep.ac.cn}
\affiliation{Theoretical Physics Division, Institute of High Energy Physics, Chinese Academy of Sciences, Beijing 100049, China}
\affiliation{School of Physical Sciences, University of Chinese Academy of Sciences, Beijing 100049, China}

\begin{abstract}
As an attractive candidate for dark matter, the \acp{PBH} in the mass range ($8\times10^{14} \sim 10^{16}$)~$\mathrm{g}$ could be detected via their Hawking radiation, including neutrinos and antineutrinos of three flavors. In this paper, we investigate the possibility to constrain the \acp{PBH} as dark matter by measuring (anti)neutrino signals at the large liquid-scintillator detector of Jiangmen Underground Neutrino Observatory (JUNO). Among six available detection channels, the inverse beta decay $\overline{\nu}^{}_e + p \to e^+ + n$ is shown to be most sensitive to the fraction $f^{}_{\rm PBH}$ of \acp{PBH} contributing to the dark matter abundance. Given the \ac{PBH} mass $M^{}_{\rm PBH} = 10^{15}~{\rm g}$, we find that JUNO will be able to place an upper bound $f^{}_{\rm PBH} \lesssim 3\times 10^{-5}$, which is 20 times better than the current best limit $f^{}_{\rm PBH} \lesssim 6\times 10^{-4}$ from Super-Kamiokande. For heavier \acp{PBH} with a lower Hawking temperature, the (anti)neutrinos become less energetic, leading to a relatively weaker bound.
\end{abstract}

\maketitle

\section{Introduction}\label{sec:intro}

The observations of gravitational waves by \ac{LIGO} \cite{Abbott:2016blz} have stimulated very active discussions on the origin of \acp{BBH}. In particular, great interest in the primordial black holes (PBHs) as the seeds for the formation of \acp{BBH} has been revived \cite{Bird:2016dcv, Wang:2016ana, Mandic:2016lcn, Clesse:2016ajp, Raidal:2017mfl, Cholis:2016xvo, Clesse:2016vqa, Cai:2018dig, Sasaki:2016jop}. 
Recent studies \cite{Sasaki:2016jop,Ali-Haimoud:2017rtz,Chen:2018czv} have shown that the \ac{PBH} scenario predicts a merger rate that is very consistent with the local observations. On the other hand, produced in the early Universe, \acp{PBH} could be a viable candidate for cold dark matter \cite{Hawking:1971ei, Carr:1974nx, Chapline:1975ojl}. The fraction of dark matter in the form of \acp{PBH} is usually defined as $f^{}_{\text{PBH}}=\Omega_{\text{PBH}}/\Omega_{\text{DM}}$, where $\Omega_{\text{PBH}}$ and $\Omega_{\text{DM}}$ denote the energy density fraction of \acp{PBH} and that of dark matter in the present Universe, respectively. Various constraints on $f^{}_{\text{PBH}}$ within a broad range of \ac{PBH} masses have been derived in the literature. See recent reviews in Refs. \cite{Carr:2020gox,Carr:2020xqk} and references therein. 

The Hawking radiation \cite{Hawking:1974sw} of the \acp{PBH} has been suggested for experimental observations and can be used to explore the intrinsic properties of \acp{PBH} \cite{Boudaud:2018hqb, Dasgupta:2019cae, 1906.04750, 0912.5297, Laha:2019ssq, Laha:2020ivk, Ballesteros:2019exr, Lunardini:2019zob}. As is well known, the black holes could emit particles near their event horizons, and the black hole evaporates faster as its mass decreases \cite{Page1976Particle, Macgibbon1990Quark, Macgibbon1991Quark}. For a \ac{PBH}, depending on its formation time, it could have a much smaller mass than ordinary stars \cite{Hawking:1971ei, Carr:1974nx}. The \acp{PBH} with masses $M_{\mathrm{PBH}}\lesssim \mathcal{O}(10^{15})~\mathrm{g}$ would have evaporated over at the present age of the Universe \cite{Hawking:1974sw}. Based on the experimental searches for the Hawking radiation, stringent constraints on $f^{}_{\text{PBH}}$ within a mass range $(10^{14}\sim 10^{17})~\mathrm{g}$ have been obtained (see, e.g., Refs. \cite{Carr:2020gox,Carr:2020xqk}, for recent reviews). For example, there exists an upper limit $f^{}_{\text{PBH}} \lesssim 10^{-7}$ for $M^{}_{\mathrm{PBH}}\simeq10^{15}\text{g}$ through an observation of the $e^\pm$ flux from Voyager \textit{1} \cite{Boudaud:2018hqb}. By observing the $\gamma$-ray lines from the $e^-$-$e^+$ annihilation and the isotropic diffuse $\gamma$-ray background, respectively, \ac{INTEGRAL} \cite{Siegert:2016ijv} and \ac{FERMI-LAT} \cite{Ackermann:2014usa} have restricted $f_{\text{PBH}}$ to be less than $10^{-3}$ for $M^{}_{\mathrm{PBH}}=10^{16}\text{g}$ \cite{Dasgupta:2019cae, 1906.04750}. Reference~\cite{Dasgupta:2019cae}, also shows that an upper limit $f^{}_{\text{PBH}}\lesssim 6 \times 10^{-4}$ for $M^{}_{\mathrm{PBH}} \simeq 10^{15}~\text{g}$ can be derived via the neutrino observation from \ac{Super-K}  \cite{Bays:2012iy}. 

As a multipurpose neutrino experiment with a 20 kton liquid-scintillator (LS) detector, \ac{JUNO} \cite{1507.05613} will be able to detect low-energy astrophysical neutrinos in addition to reactor electron-antineutrinos. The primary reaction channels for astrophysical neutrinos at \ac{JUNO} include the inverse beta decay $\overline{\nu}^{}_e + p \to e^+ + n$ (IBD), the elastic neutrino-proton scattering $\nu + p \to \nu + p$ ($p$ES), the elastic neutrino-electron scattering $\nu^{}_e + e^- \to \nu^{}_e + e^-$ ($e$ES), and the neutrino-nucleus (i.e., ${^{12}}{\rm C}$) reactions $\nu + {^{12}}{\rm C} \to \nu +  {^{12}}{\rm C}^*$, $\nu^{}_e +  {^{12}}{\rm C} \to  e^- + {^{12}}{\rm N}$, and $\overline{\nu}^{}_e + {^{12}}{\rm C} \to e^+ + {^{12}}{\rm B}$. With a perfect neutron-tagging efficiency of the LS detector, \ac{JUNO} is expected to have a great potential to discover the \ac{DSNB} \cite{1507.05613}. As mentioned above, the upper limit on the \ac{DSNB} flux from \ac{Super-K} has been used to constrain $f^{}_{\text{PBH}}$ for small-mass \acp{PBH} \cite{Dasgupta:2019cae}. Therefore, we are very motivated to investigate how restrictive the bound on $f^{}_{\rm PBH}$ from \ac{JUNO} will be. Compared to \ac{Super-K}, we expect that \ac{JUNO} could improve the constraint on $f^{}_{\mathrm{PBH}}$ due to its more powerful neutron tagging, lower energy threshold, and better energy resolution. A similar analysis can be performed for the future Gadolinium-doping upgrade of \ac{Super-K}. 

The remaining part of this paper is organized as follows. In Sec.~\ref{sec:flux}, we calculate the neutrino fluxes from \acp{PBH} in the Galactic and extragalactic dark matter halos. Then, Sec.~\ref{sec:spectrum} is devoted to the event spectra for all six neutrino detection channels at \ac{JUNO} and the estimation of the relevant backgrounds. In Sec.~\ref{sec:sensitivity}, by comparing between the signals and backgrounds, we draw the upper limit on $f^{}_{\rm PBH}$ from JUNO. Our main conclusions are finally summarized in Sec.~\ref{sec:con}.

\section{Neutrino fluxes from \acp{PBH}}\label{sec:flux}

Generally speaking, there are two different contributions, i.e., the primary and secondary components, to the neutrino fluxes from the evaporating \acp{PBH}. The former arises directly from the Hawking radiation, while the latter stems from the decays of the secondary particles produced in the Hawking radiation \cite{0912.5297,arbey_blackhawk:_2019}. For an evaporating \ac{PBH}, the number of (anti)neutrinos ($N$) in unit energy ($E$) and time ($t$) is given by 
\begin{equation}\label{eq:tot}
\frac{{\rm d}^2N}{{\rm d}E {\rm d}t} = \frac{{\rm d}^2N}{{\rm d}E{\rm d}t} \Bigg|_{\text{pri}} + \frac{{\rm d}^2N}{{\rm d}E {\rm d}t}\Bigg|_{\text{sec}} \; ,
\end{equation}
where the first and second terms on the right-hand side denote the primary and secondary components, respectively. In our calculations, both components are evaluated by using \texttt{BLACKHAWK} \cite{arbey_blackhawk:_2019}. It is worthwhile to notice that the spins of \acp{PBH} in the present work are assumed to be negligible, as predicted by a class of theories for the \ac{PBH} production \cite{Chiba_2017,Mirbabayi:2019uph}.

Then the differential fluxes of (anti)neutrinos radiated from \acp{PBH} can be calculated by taking into account the distribution and cosmological evolution. In units of $\text{cm}^{-2}~\text{s}^{-1}~\text{MeV}^{-1}$, they can be decomposed as follows
\begin{equation}\label{eq:dfde}
\frac{{\rm d}F}{{\rm d}E} = \frac{{\rm d}F^{}_{\text{Gal}}}{{\rm d}E} + \frac{{\rm d}F^{}_{\text{EG}}}{{\rm d}E} \; ,
\end{equation}
where the first and second terms at the right-hand side are contributed by \acp{PBH} located in the Galactic and extragalactic dark halos, respectively. 

For the extragalactic \acp{PBH}, the differential (anti)neutrino flux can be written as \cite{Dasgupta:2019cae,0912.5297}
\begin{equation}\label{eq:EG}
    \begin{split}
    \frac{{\rm d}F^{}_{\text{EG}}}{{\rm d}E} = \int_{t_{\text{min}}}^{t_{\text{max}}} & {\rm d}t \ [1 + z(t)] \frac{f^{}_{\text{PBH}} \rho^{}_{\text{DM}}}{M^{}_{\text{PBH}}}\\
    & \times \frac{{\rm d}^2N}{{\rm d}E_{\rm s} {\rm d}t} \bigg|_{E^{}_{\rm s} = [1 + z(t)]E} \; .
    \end{split}
\end{equation}
According to the \emph{Planck} 2018 results \cite{Aghanim:2018eyx}, we set the average energy density of dark matter in the present Universe to be $\rho^{}_{\text{DM}} = 2.35 \times 10^{-30}~\text{g}~\text{cm}^{-3}$. The (anti)neutrino energy at the source has been denoted as $E^{}_{\rm s}$, while that in the observer's frame has been denoted as $E$. The connection between them is given by $E^{}_{\rm s} = [1 + z(t)] E$, where the redshift $z(t)$ is a function of the cosmic time $t$ and encodes the cosmological evolution. For the lower and upper limits of the integration over the cosmic time in Eq.~(\ref{eq:EG}), we take the following values. 
First, the (anti)neutrinos emitted from the \acp{PBH} in the very early time will be significantly redshifted. This causes that these (anti)neutrinos have extremely low energies today and cannot be detected.
There is an exception for high-energy (anti)neutrinos, whose fluxes, however, are highly suppressed for the \ac{PBH} masses of our interest. Therefore, we fix $t^{}_{\text{min}}=10^{11}~\text{s}$ close to the epoch of radiation-matter equality and numerically confirm that changing $t^{}_{\rm min}$ to be smaller has essentially no impact on the final results. Second, we choose $t^{}_{\text{max}}=\text{min}\{\tau^{}_0,\tau^{}_{\text{PBH}}\}$, where $\tau^{}_0$ is the age of the Universe, $\tau^{}_{\text{PBH}}$ is the lifetime of \acp{PBH}, and the function $\text{min}\{x^{}_1,x^{}_2\}$ singles out the smaller one from $x^{}_1$ and $x^{}_2$. 

For the \acp{PBH} in the Galactic dark halo, the differential (anti)neutrino flux can be calculated as \cite{Dasgupta:2019cae}
\begin{equation}\label{eq:Gal}
\frac{{\rm d}F^{}_{\mathrm{Gal}}}{{\rm d}E} = \int \frac{{\rm d}\Omega}{4\pi} \int_0^{l^{}_{\text{max}}} {\rm d}l~ \frac{{\rm d}^2N}{{\rm d}E {\rm d}t} \frac{f^{}_{\text{PBH}}~\rho^{}_{\text{Gal}}[r(l,\psi)]}{M^{}_{\text{PBH}}} \; ,
\end{equation}
where $r(l, \psi) \equiv \sqrt{r^2_\odot - 2l r^{}_\odot \cos\psi + l^2}$ is the galactocentric distance calculated from the distance of the Earth to the Galactic center $r^{}_\odot = 8.5~{\rm kpc}$ and the line-of-sight distance $l$ to the \ac{PBH}, with $\psi$ being the angle between these two directions. In addition, the angular integration is defined as $\int {\rm d}\Omega = \int_0^{2\pi} {\rm d}\phi \int_0^\pi {\rm d}\psi \sin\psi$ with $\phi$ being the azimuthal angle, and $\rho^{}_{\text{MW}}(r)$ is the local energy density of dark matter. The maximal value of $l$ is determined by $l^{}_{\text{max}}=({r^2_{\rm h} - r^2_\odot \sin^2\psi})^{1/2} + r^{}_\odot \cos \psi$ with the halo radius $r^{}_{\rm h} = 200~{\rm kpc}$. For illustration, we implement the \ac{NFW} profile \cite{ng_resolving_2014}
\begin{eqnarray}\label{eq:NFW}
    \rho^{}_{\text{NFW}}(r) = 0.4\times\left(\frac{8.5}{r}\right)\left(\frac{1+{8.5}/{20}}{1+{r}/{20}}\right)^2 \ ,
\end{eqnarray}
which is in units of $\text{GeV}\cdot\text{cm}^{-3}$, and the distance $r$ is in units of $\text{kpc}$. 

\begin{figure}
    \includegraphics[width = 0.45\textwidth]{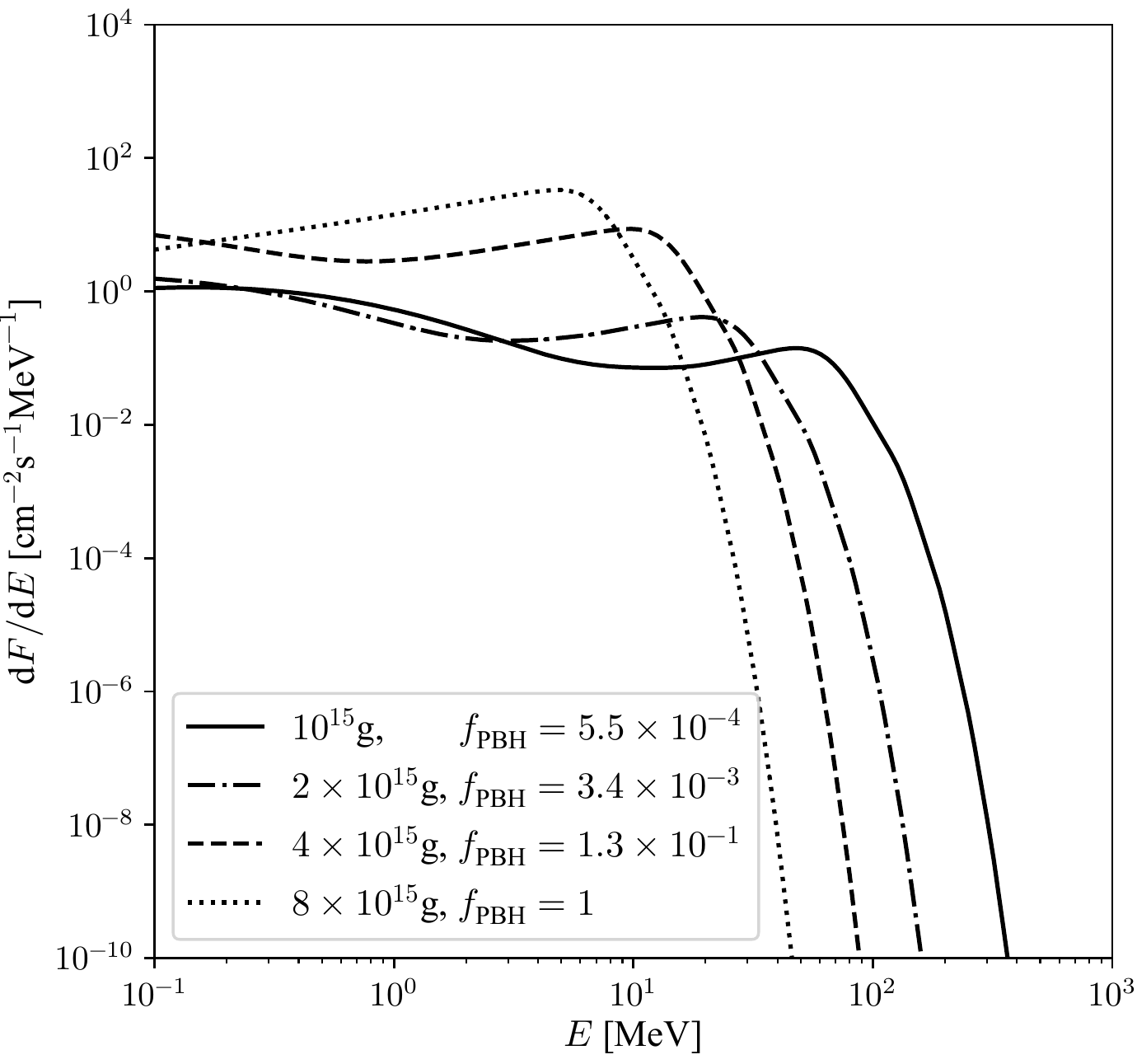}
    \caption{The differential (anti)neutrino flux from \acp{PBH} for $M^{}_{\rm PBH} = 10^{15}~{\rm g}$ (solid curve), $2\times 10^{15}~{\rm g}$ (dot-dashed curve), $4\times 10^{15}~{\rm g}$ (dashed curve) and $8\times 10^{15}~{\rm g}$ (dotted curve), where the monochromatic mass function of \acp{PBH} is assumed and the upper limit on $f^{}_{\rm PBH}$ from \ac{Super-K} \cite{Dasgupta:2019cae} is input. The horizontal axis represents the (anti)neutrino energy in units of $\mathrm{MeV}$ at the detector, while the vertical one denotes the differential (anti)neutrino flux in units of $\text{cm}^{-2}~\text{s}^{-1}~\text{MeV}^{-1}$.}\label{fig:different_f}  
\end{figure}
In Fig.~\ref{fig:different_f}, we show the differential fluxes of $\nu^{}_{e}$ emitted from \acp{PBH} for four benchmark masses within the range $(10^{15} \sim 10^{16})~\mathrm{g}$. The results for neutrinos and antineutrinos of other flavors are quite similar, as a consequence of the thermal Hawking radiation. In the numerical calculations, we have assumed a monochromatic mass function of \acp{PBH} and taken the existing upper limit of $f^{}_{\rm PBH}$ from \ac{Super-K} \cite{Dasgupta:2019cae}. As one can observe from Fig.~\ref{fig:different_f}, neutrinos and antineutrinos from smaller-mass \acp{PBH} have higher energies, mainly due to the higher Hawking temperature, while the magnitude of the fluxes is limited by the existing bound on $f^{}_{\rm PBH}$. 

Note that we have assumed neutrinos to be Majorana particles, and will calculate the event spectra in the next section in this case as well. The difference between the cases of Majorana and Dirac neutrinos will be clarified in Sec.~\ref{sec:sensitivity}.

\section{Event Spectra and Backgrounds}\label{sec:spectrum}

Given the (anti)neutrino fluxes radiated by \acp{PBH}, we can figure out the event spectra for six available detection channels in the LS detector of \ac{JUNO}. In this section, we present the results of the event spectra and estimate the relevant backgrounds at JUNO. 

\subsection{IBD channel}

For the IBD channel $\overline{\nu}^{}_e + p \to e^+ + n$, both the final-state $e^+$ and $n$ can be perfectly observed in the LS detector. At JUNO, the event spectrum of IBD is given by \cite{1712.06985}
\begin{equation}\label{eq:IBD_event_spectrum}
\frac{{\rm d}N}{{\rm d}E_0} = N^{}_p T \int_{E^{\rm thr}_{\rm IBD}}^\infty \sigma^{\text{IBD}}_{{\bar{\nu}_e}}(E)\frac{\rm d F}{\rm d E}\Big{|}_{\bar{\nu}_e} \mathcal{G}(E^{}_0;E^{}_{\rm v},\delta^{}_E) {\rm d}E \; ,
\end{equation}
where the energy threshold for the IBD reaction is $E^{\rm thr}_{\rm IBD} = 1.8~{\rm MeV}$, $E^{}_0$ is the observed energy, $\mathcal{G}(E_0;E_{\rm v},\delta_E)$ is a Gaussian function of $E^{}_0$ with an expectation value $E^{}_{\rm v}$ and the standard deviation $\delta^{}_E = 3\%/\sqrt{E^{}_0/\text{MeV}}$, i.e., the energy resolution of JUNO \cite{1507.05613,1712.06985}. The visible energy in the detector $E^{}_{\rm v} = m^{}_e + E^{}_{e^+}$ arises from the annihilation of the final-state positrons with ambient electrons, where $m^{}_e = 0.511~{\rm MeV}$ is the electron mass. The IBD cross section $\sigma^{\text{IBD}}_{{\bar{\nu}_e}}$ and the positron energy $E^{}_{e^+}$ depend on the antineutrino energy $E$ in the differential flux ${\rm d}F/{\rm d}E$ defined in Eq.~(\ref{eq:dfde}). We obtain the dependency relation in Table~1 of Ref.~\cite{Strumia:2003zx}. The total number $N^{}_p$ of target protons in the LS and the effective running time $T$ in \citeeq{eq:IBD_event_spectrum} can be found in Ref.~\cite{1712.06985}. 

With the differential fluxes of $\overline{\nu}^{}_e$ in \citefig{fig:different_f}, we compute the IBD event spectra by using \citeeq{eq:IBD_event_spectrum} and present the final results in \citefig{fig:IBD_tot}, where different curves correspond to four benchmark \ac{PBH} masses in \citefig{fig:different_f}. In our calculations, the fiducial mass of the JUNO detector is taken to be $20$ kton and the operation time is set to $T = 10~{\rm yr}$. In \citefig{fig:IBD_tot}, we can observe that the peak of the event spectrum moves toward lower energies as the \ac{PBH} mass increases. It should be noticed that the peak rate is limited by the existing upper bound on $f^{}_{\text{PBH}}$.
\begin{figure}
    \includegraphics[width = 0.45\textwidth]{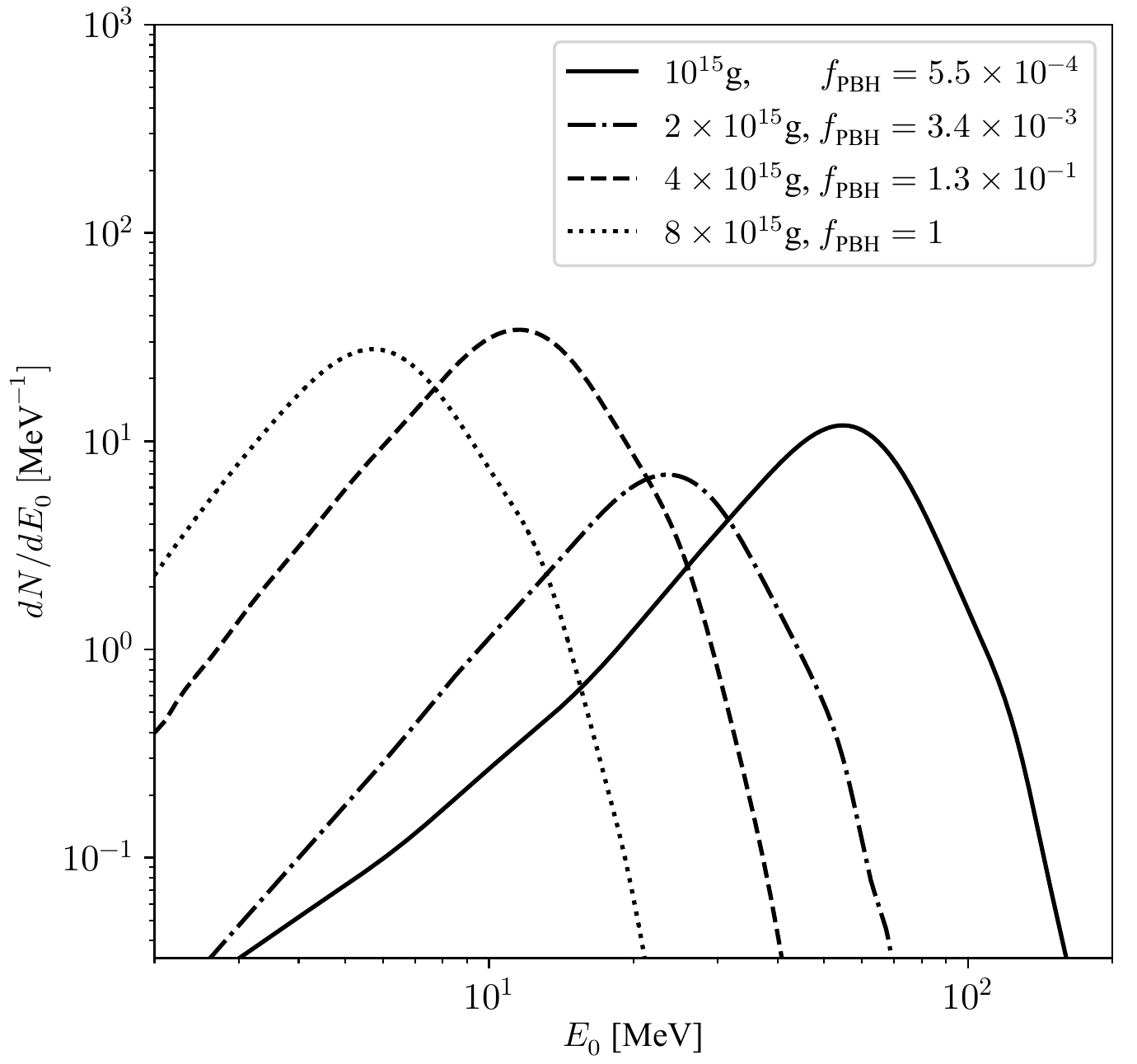}
    \caption{The IBD event spectra for \acp{PBH} of four benchmark masses as in \citefig{fig:different_f}. For the \ac{LS} detector, we have assumed the fiducial mass of $20$ kton and the effective running time of $10$~years. The horizontal axis refers to the observed energy in units of $\text{MeV}$, while the vertical one to the event spectrum in units of $\text{MeV}^{-1}$. }\label{fig:IBD_tot}
\end{figure}

The main backgrounds for the IBD channel can be divided into four categories. The first one is the irreducible background from reactor antineutrinos, which dominate over others in the energy range below $12~\text{MeV}$ \cite{1507.05613,Li:2020gaz}. 
From \citefig{fig:IBD_tot}, we notice that when the \ac{PBH} mass is $8 \times 10^{15}~\text{g}$ or larger, the peaks of the event spectra shift to the region $E^{}_0 \lesssim 10~\text{MeV}$. Consequently, for $ M^{}_{\rm PBH} \gtrsim 8\times10^{15}\text{g}$, the IBD signal will be severely contaminated by the reactor antineutrino background. For this reason, we mainly focus on the \acp{PBH} with smaller masses or set an energy cut $E^{}_0 \gtrsim 12~{\rm MeV}$ to get rid of this background. The second one is \ac{DSNB}, which is one the primary goals of \ac{Super-K} and \ac{JUNO} \cite{1507.05613}. For illustration, we take into account the \ac{DSNB} $\overline{\nu}^{}_e$ flux with an average energy $\langle E^{}_{\bar{\nu}_e}\rangle = 15~\text{MeV}$. 
The third type of background is composed of the atmospheric neutrino charged-current (atm.CC) background and neutral-current (atm.NC) background. The former is caused by the IBDs of atmospheric neutrino $\bar{\nu}_e$, while the latter by the \ac{NC} reactions of high-energy atmospheric (anti)neutrinos in the \ac{LS} \cite{1507.05613,Collaboration:2011jza}.
Both kinds of event spectra have been studied in Ref.~\cite{1507.05613}. The fourth background is the fast neutrons (FNs), which are produced by the cosmic high-energy muons arriving at the detector and induce the IBD-like events \cite{1507.05613}. For the last two categories of backgrounds, the dedicated method of \ac{PSD} has to be utilized, as we shall explain below.
\begin{table*}[!t]
    \centering
    \begin{tabular}{|m{4cm}<{\centering}|m{1.25cm}<{\centering}|m{1.25cm}<{\centering}|m{1.25cm}<{\centering}|m{1.25cm}<{\centering}|m{1.25cm}<{\centering}|m{1.25cm}<{\centering}|m{1.25cm}<{\centering}|m{1.25cm}<{\centering}|}
        \hline
        \hline
        $M^{}_{\rm PBH}$[g] & \multicolumn{2}{c|}{$10^{15}$} & \multicolumn{2}{c|}{$2\times10^{15}$} & \multicolumn{2}{c|}{$4\times10^{15}$} & \multicolumn{2}{c|}{$8\times10^{15}$} \\
        \hline
        $f_\text{PBH}$ (Super-K constraints) & \multicolumn{2}{c|}{$5.5\times10^{-4}$} & \multicolumn{2}{c|}{$3.4\times10^{-3}$} & \multicolumn{2}{c|}{$1.3\times10^{-1}$} & \multicolumn{2}{c|}{$1$} \\
        \hline
        Energy window [MeV] & \multicolumn{2}{c|}{$[12,130]$} & \multicolumn{2}{c|}{$[12,80]$} & \multicolumn{2}{c|}{$[12,40]$} & \multicolumn{2}{c|}{$[12,20]$} \\
        \hline
        PSD procedure & Before & After & Before & After & Before & After & Before & After \\
        \hline
        PBH signal & $541.88$ & $270.94$ & $137.59$ & $68.80$ & $203.07$ & $101.54$ & $7.79$ & $3.90$ \\
        \hline
        Atm.NC & $1333.34$ & $14.67$ & $1105.10$ & $12.16$ & $746.71$ & $8.21$ & $348.24$ & $3.83$ \\
        \hline
        Atm.CC & $33.31$ & $16.66$ & $17.36$ & $8.68$ & $3.79$ & $1.90$ & $0.46$ & $0.23$ \\
        \hline
        DSNB & $17.72$ & $8.86$ & $17.72$ & $8.86$ & $17.72$ & $8.86$ & $12.20$ & $6.10$ \\
        \hline 
        FN & $83.78$ & $1.42$ & $48.28$ & $0.82$ & $19.88$ & $0.34$ & $5.68$ & $0.10$ \\
        \hline
        Total backgrounds & $1468.15$ & $41.61$ & $1188.46$ & $30.52$ & $788.10$ & $19.31$ & $366.58$ & $10.26$\\
        \hline
        $f_\text{PBH}$ (JUNO constraints) & \multicolumn{2}{c|}{$2.1\times10^{-5}$} & \multicolumn{2}{c|}{$4.4\times10^{-4}$} & \multicolumn{2}{c|}{$9.0\times10^{-3}$} & \multicolumn{2}{c|}{$1$} \\
        \hline
        \hline
    \end{tabular}
    \caption{Summary of the upper limits of the \ac{IBD} event rates according to the \ac{Super-K} constrain, the background rates estimation according to the corresponding mass of \ac{PBH} energy windows from \ac{JUNO} detector, the comparison of before and after \ac{PSD} cut application, and the $f_\text{PBH}$ constrain estimation from \ac{JUNO} backgrounds. The event rates are given in units of counts per $10$ year per $20$ kton.
        }\label{tab:IBD_PSD}
\end{table*}

Similar to the \ac{DSNB} searches in the \ac{LS} detectors, the \ac{PSD} approach is critically important to enhance the \acp{SNR}. Both signals and backgrounds in the \ac{LS} will be finally converted into light, which can be observed by the photomultiplier tubes. For the light signals generated in different processes, the probability density functions of the photon emission time can be described as \cite{Mollenberg:2014pwa}
\begin{equation}\label{eq:PSD}
    F(t) = \sum_i\frac{N^{}_i}{\tau^{}_i}e^{-t/\tau^{}_i} \; ,
\end{equation}
where $i$ refers to the fast, slow and slower components, and $N^{}_i$ and $\tau^{}_i$ denote the fraction and time constant of $i$-th component, respectively. Then one can in principle identify different particles by the corresponding distribution functions $F(t)$. To make use of the tail-to-total ratio \cite{Mollenberg:2014pwa}, we define the working parameter \ac{PSD} as
\begin{equation}\label{eq:tail-to-total}
    \text{PSD} = \frac{\displaystyle \int_{t^{}_\text{cut}}^\infty F(t) {\rm d}t}{\displaystyle \int_0^\infty F(t) {\rm d}t} \; ,
\end{equation}
where $t^{}_\text{cut}$ denotes a cutoff time. For the signal and one kind of the relevant backgrounds, one can properly choose the cutoff time in order to reduce the number of background events more than that of the signal events. 

For \ac{JUNO}, Ref. \cite{1507.05613} has estimated the \ac{PSD} efficiency, which is the fraction of events surviving the \ac{PSD} procedure, $\epsilon^{}_\nu = 0.5$, $\epsilon^{}_{\text{NC}} = 0.011$, and $\epsilon^{}_{\text{FN}} = 0.017$.

In Table~\ref{tab:IBD_PSD}, we summarize the total number of IBD events from \acp{PBH} and that of relevant backgrounds, where both the results before and after applying the \ac{PSD} approach have been shown for comparison. The IBD events from \acp{PBH} are calculated as in \citefig{fig:IBD_tot}, and the backgrounds are categorized as in previous discussions. Comparing the total backgrounds before and after applying the \ac{PSD} approach, we find a significant reduction of the backgrounds. One can notice that the dominant background is due to atm.NC, and for $M^{}_{\rm PBH} > 10^{15}~\text{g}$, the \acp{SNR} without the \ac{PSD} procedure are too small for any realistic observations.

\subsection{$e$ES channel}

For the $e$ES channel $\nu + e^- \to \nu + e^-$, all species of neutrinos and antineutrinos could contribute and the event spectrum can be calculated as \cite{1712.06985}
\begin{equation}\label{eq:eES_event_spectrum}
    \begin{split}
        \frac{{\rm d}N}{{\rm d}E^{}_0} = & N^{}_e T \sum_\alpha \int_0^\infty {\rm d}T^{}_e \mathcal{G}(E^{}_0;T^{}_e,\delta^{}_E) \\
        & \times \int_{E_e^{\text{min}}}^\infty \frac{{\rm d}F}{{\rm d}E} \bigg|_\alpha \frac{{\rm d}\sigma^{e\text{ES}}_{\nu_\alpha}(E)}{{\rm d} T^{}_e} {\rm d}E \; ,
    \end{split}
\end{equation}
where the $e$ES cross section is $\sigma^{e\text{ES}}_{\nu^{}_\alpha}$ with $\alpha$ standing for neutrinos and antineutrinos of three flavors, and $T^{}_e$ is the electron kinetic energy. The minimal neutrino energy to produce a recoil electron energy $T^{}_e$ is determined by $E_e^{\text{min}} \simeq T^{}_e/2 + \sqrt{T^{}_e(T^{}_e + 2m^{}_e)}/2$. In addition, we should set $N^{}_e = N^{}_p$ due to the electric neutrality of the \ac{LS} materials.

\begin{figure}
    \includegraphics[width = 0.45\textwidth]{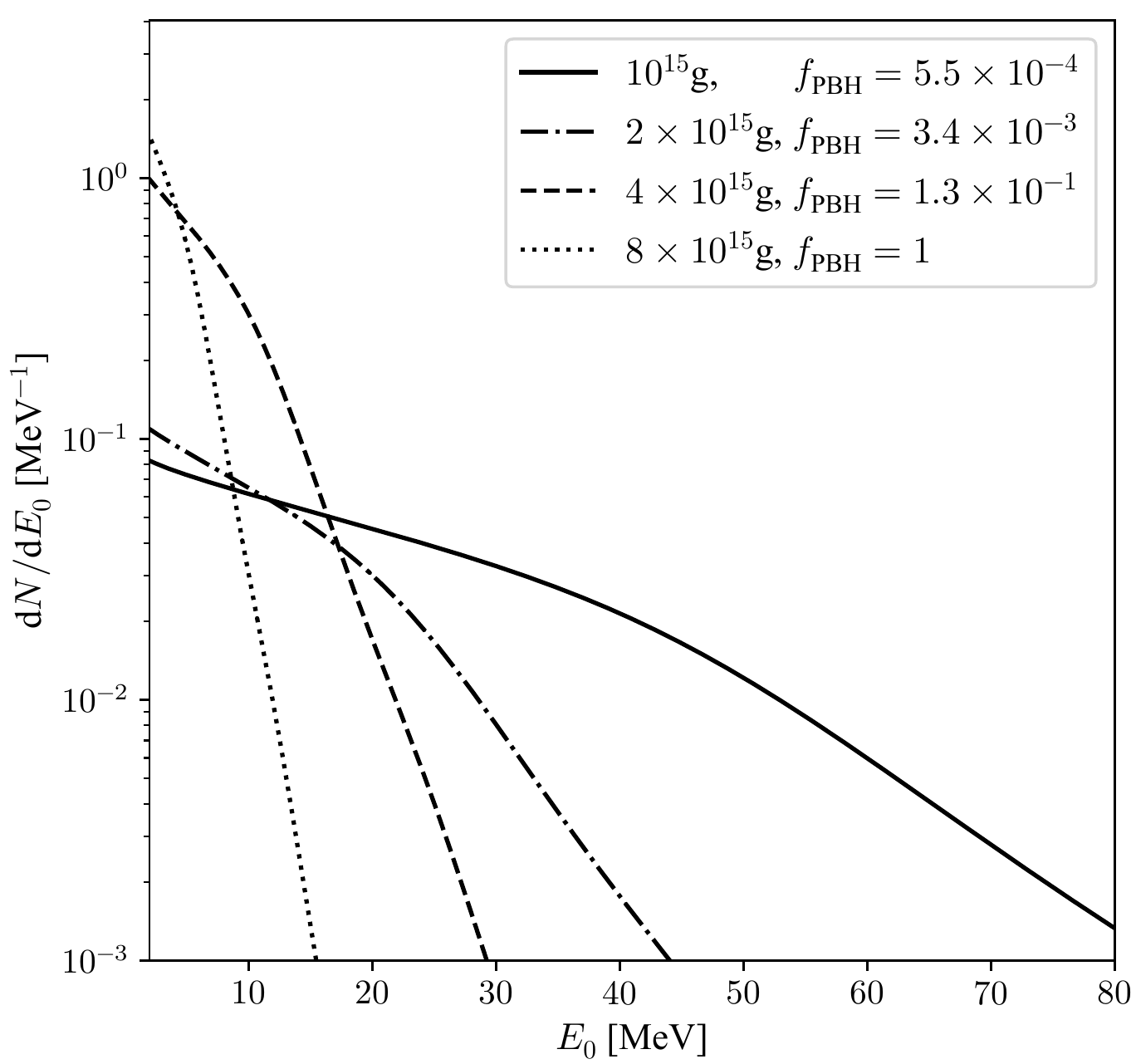}
    \caption{The $e$ES event spectra induced by \acp{PBH} for four benchmark masses corresponding to those in \citefig{fig:IBD_tot}, where the effective running time of $10$ years and the fiducial \ac{LS} mass of $20$ kton have been assumed.}\label{fig:eES_tot}
\end{figure}
The $e$ES event spectra from \acp{PBH} are shown in \citefig{fig:eES_tot}, where four benchmark masses for the \acp{PBH} corresponding to those in \citefig{fig:IBD_tot} have been taken and the effective running time of 10 years and the fiducial \ac{LS} mass of 20 kton have been assumed. As one can observe from \citefig{fig:eES_tot}, the maximal rate reaches $1.5~{\rm MeV}^{-1}$ for $M^{}_{\rm PBH} = 8\times 10^{15}~{\rm g}$ at the low-energy end. However, for the electron recoil events, there are numerous backgrounds from the $^8$B solar neutrino and other three types of cosmogenic isotopes (i.e., $^{11}$C, $^{10}$C, and $^{11}$Be). According to Ref.~\cite{1507.05613}, the total background events are estimated to $10^4\ \text{MeV}^{-1}$ per $10$ years per $20$ kton at \ac{JUNO}. Compared with this background rate, the $e$ES event rates shown in \citefig{fig:eES_tot} are much smaller. Therefore, it seems impossible to observe the $e$ES signals from \acp{PBH} at the \ac{LS} detectors. 

\subsection{$p$ES channel}

As the advantages of the low energy threshold of the \ac{LS} detectors, the $p$ES channel $\nu + p \to \nu + p$ can also be used to observe astrophysical neutrinos. The $p$ES event spectrum is computed as follows \cite{1712.06985}
\begin{equation}\label{eq:pES_event_spectrum}
    \begin{split}
        \frac{{\rm d}N}{{\rm d}E^{}_0} = & N^{}_p T \int {\rm d} T_p^\prime \mathcal{G}(E^{}_0;T_p^\prime,\delta^{}_E) \\
        & \times \sum_\alpha \frac{{\rm d}T^{}_p}{{\rm d}T_p^\prime} \int_{E_p^{\text{min}}}^\infty \frac{{\rm d}F}{{\rm d}E}\bigg|_\alpha \frac{{\rm d}\sigma^{p\text{ES}}_{\nu^{}_\alpha}(E)}{{\rm d}T^{}_p} {\rm d}E \; ,
    \end{split}
\end{equation}
where ${\rm d}\sigma^{p{\rm ES}}_{\nu^{}_\alpha}/{\rm d}T^{}_p$ denotes the differential cross section with $T^{}_p$ being the proton recoil energy. The minimal neutrino energy required to produce the final-state proton with a recoil energy of $T^{}_p$ is approximately determined by $E_p^{\text{min}} = (T^{}_p m^{}_p/2)^{1/2}$ with $m^{}_p = 938.27~{\rm MeV}$ being the proton mass. However, the proton recoil energy will be quenched in the \ac{LS} and observed as $T_p^\prime$, which is related to $T^{}_p$ by the Birks law \cite{1712.06985,Beacom:2002hs}. The quenching effects on recoiled protons in the \ac{LS} have been shown numerically in Fig.~30 of Ref. \cite{1507.05613} and will be taken into account in our calculations. 

Since the recoil energies of protons are small and the observed energies after quenching effects become even smaller, we consider only the (anti)neutrinos from the \acp{PBH} with a mass of $M^{}_{\rm PBH} = 10^{15}~{\rm g}$, for which the (anti)neutrino energies are much higher than those for larger \ac{PBH} masses. The final result has been presented in \citefig{fig:1e15_pES}, from which one can see that the observed energy turns out to be located in the range of $(0.2 \sim 1.8)~\text{MeV}$. Below $1~\text{MeV}$, the background is mainly caused by the radioactive decays of the \ac{LS} materials and surroundings, whose rate has been estimated to be about $80$ events per $10$ sec \cite{1507.05613}. In contrast, the $p$ES event rate produced by \acp{PBH} is small, as one can observe from \citefig{fig:1e15_pES}. It should be noticed that the operation time for the \ac{LS} in \citefig{fig:1e15_pES} has been taken to be 10 years, together with the fiducial \ac{LS} mass of $20$ kton. For the observed energy higher than $1\text{MeV}$, the rate for possible backgrounds has not been given in Ref.~\cite{1507.05613}, but the signal rate is highly suppressed, rendering the realistic observation to be difficult.
\begin{figure}
    \includegraphics[width = 0.45\textwidth]{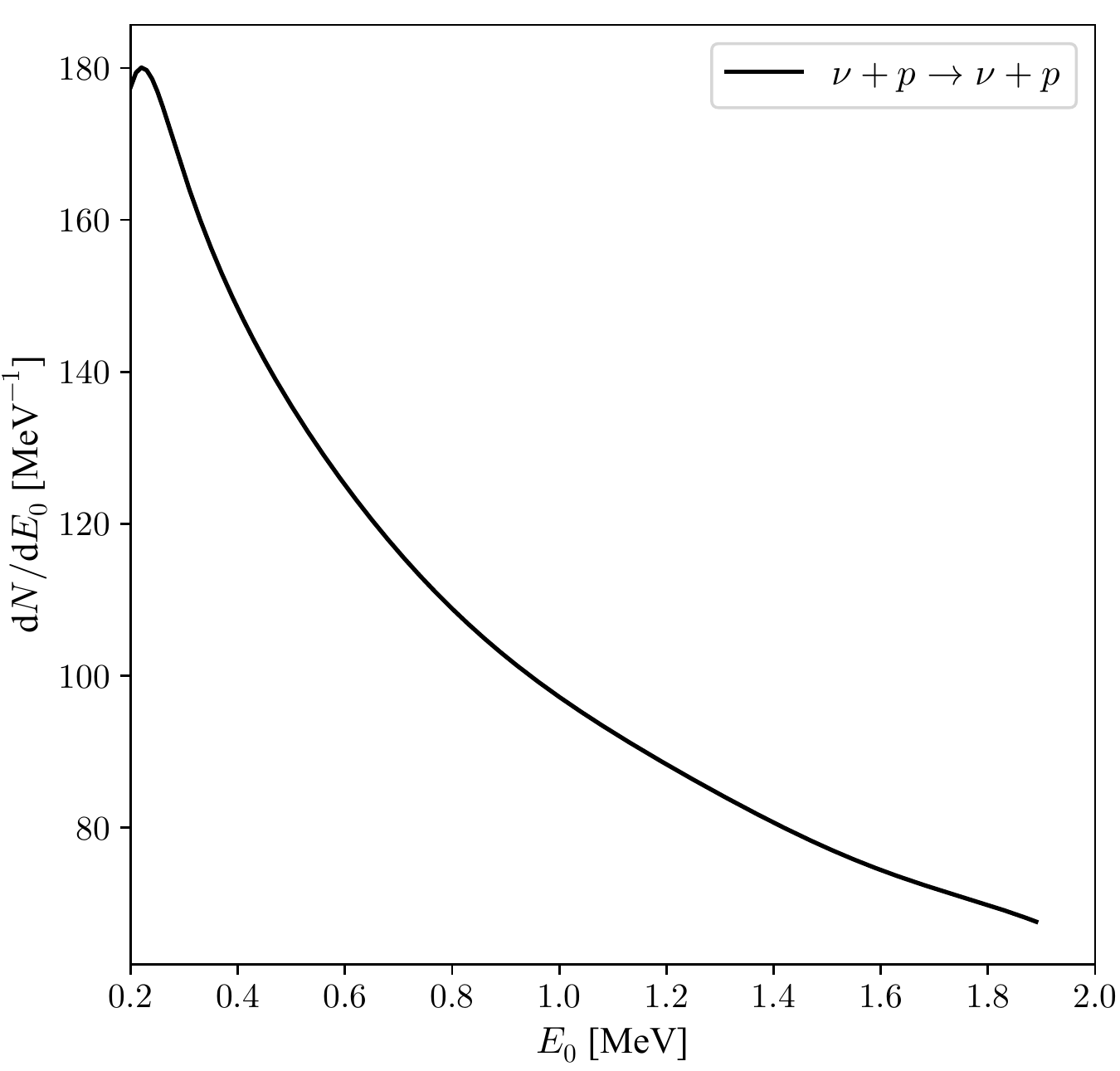}
    \caption{The $p$ES event spectrum induced by the (anti)neutrinos from the \acp{PBH} of $M^{}_{\rm PBH} = 10^{15}~\text{g}$, $f^{}_{\rm PBH}=5.5\times 10^{-4}$, where the effective running time of $10$ years and the fiducial \ac{LS} mass of $20$ kton have been assumed.}\label{fig:1e15_pES}
\end{figure}

\subsection{$^{12}$C channel}

There are three different reaction channels for the interaction between astrophysical neutrinos with the $^{12}$C nuclei in the \ac{LS} detectors. The first one is the \ac{NC} reaction $\nu + {^{12}}{\rm C} \to \nu + {^{12}}{\rm C}^*$, where the final-state nucleus resides in the excited state $(15.1~\text{MeV}, 1^+)$ and its deexcitation leads to a gamma ray of $15.1~{\rm MeV}$ that can be registered in the detector. The \ac{NC} event spectrum is given by
\begin{equation}\label{eq:NC_event_spectrum}
    \begin{split}
        \frac{{\rm d}N}{{\rm d}E^{}_0} = & N^{}_\text{C} T \sum_\alpha \int {\rm d}E \sigma^{\text{NC}}_{\nu^{}_\alpha} (E) \frac{{\rm d}F}{{\rm d}E}\bigg|_\alpha \\  & \times\mathcal{G}(E^{}_0;15.1~\text{MeV}, \delta^{}_E) \; ,  
    \end{split}
\end{equation}
where $N^{}_{\rm C} = 3 N^{}_e/23$ is the total number of $^{12}$C nuclei in the \ac{LS} target \cite{Jia:2017oar} and $\sigma^\text{NC}_{\nu^{}_\alpha}$ denotes the total cross section. The other two are \ac{CC} reactions, namely, $\nu^{}_e + {^{12}}\text{C} \to e^- + {^{12}}\text{N}$ and $\overline{\nu}^{}_e + {^{12}}\text{C} \to e^+ + {^{12}}\text{B}$. The \ac{CC} event spectrum is found to be
\begin{equation}\label{eq:CC_event_spectrum}
    \begin{split}
        \frac{{\rm d}N}{{\rm d}E^{}_0} = N^{}_\text{C} T \int {\rm d}E & \sigma^\text{CC}_{\nu^{}_e(\overline{\nu}^{}_e)}(E) \frac{{\rm d}F}{{\rm d}E}\bigg|_{\nu^{}_e (\overline{\nu}^{}_e)} \\
        & \times \mathcal{G}(E^{}_0;T^{}_e,\delta^{}_E) \; ,
    \end{split}
\end{equation}
where $T^{}_e$ is the recoil energy of the final-state electron (or positron) and $\sigma^\text{CC}_{\nu^{}_e}$ and $\sigma^{\rm CC}_{\overline{\nu}^{}_e}$ denotes respectively the cross section of $\nu^{}_e$-${^{12}}{\rm C}$ and $\overline{\nu}^{}_e$-${^{12}}{\rm C}$ reaction. Moreover, we have $T^{}_e = E^{}_{\nu^{}_e} - 16.827~\text{MeV}$ for the $^{12}\text{C}(\nu^{}_e,e^-)^{12}\text{N}$ channel while $T^{}_e = E^{}_{\overline{\nu}^{}_e} - 13.880~\text{MeV}$ for the $^{12}\text{C}(\overline{\nu}^{}_e,e^+)^{12}\text{B}$ channel \cite{Fukugita:1988hg}. The cross sections of all three neutrino-carbon reaction channels can be found in Table~1 of Ref.~\cite{Fukugita:1988hg}. 
\begin{figure}
    \includegraphics[width = 0.45\textwidth]{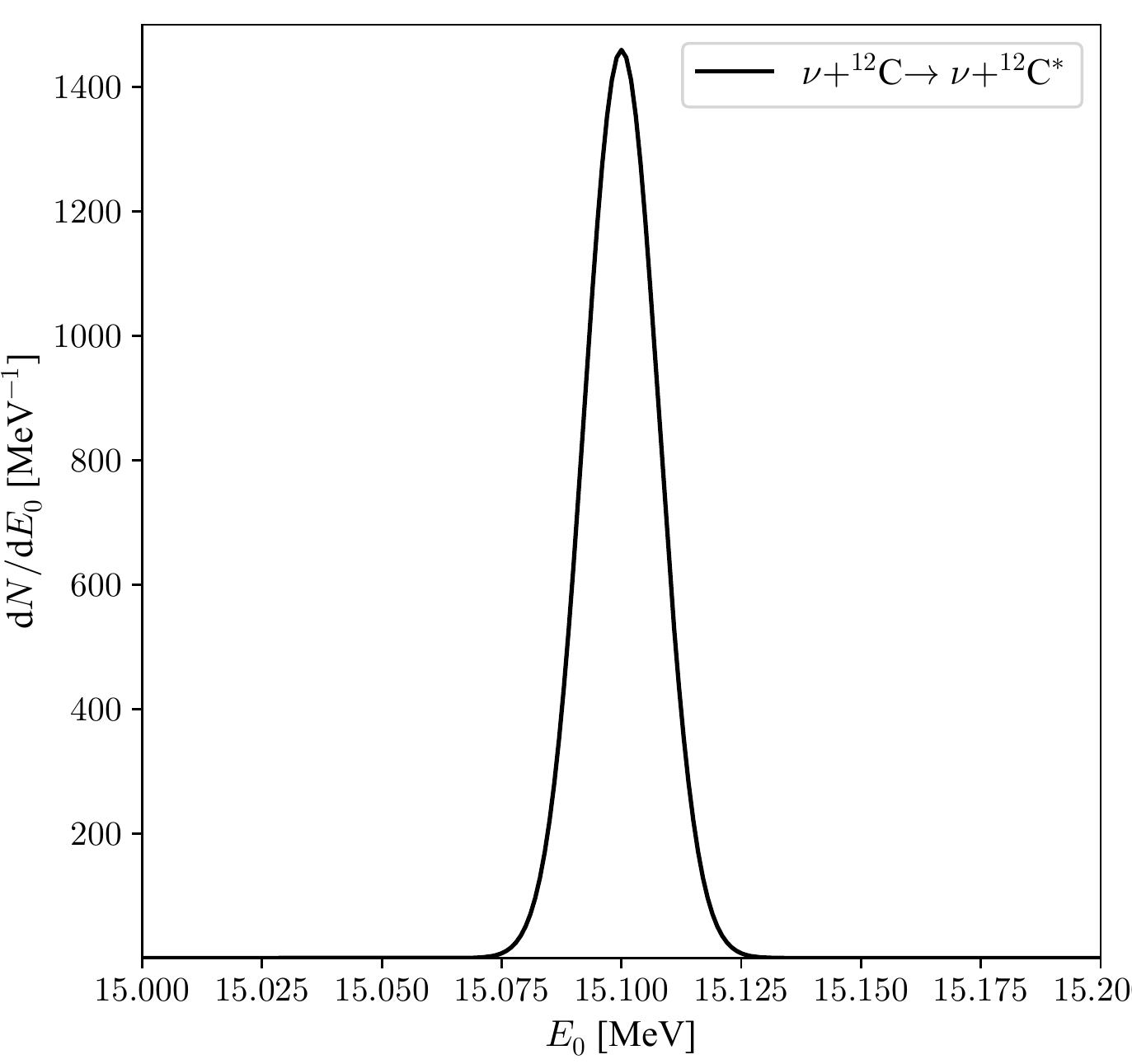}
    \caption{The event spectrum of the \ac{NC} neutrino-$^{12}$C reaction for the \acp{PBH} of $M^{}_{\rm PBH} = 10^{15}~\text{g}$, $f^{}_{\rm PBH}=5.5\times 10^{-4}$, where the effective running time of $10$ years and the fiducial \ac{LS} mass of $20$ kton are taken.}\label{fig:1e15_C12_NC}
\end{figure}
\begin{figure}
    \includegraphics[width = 0.45\textwidth]{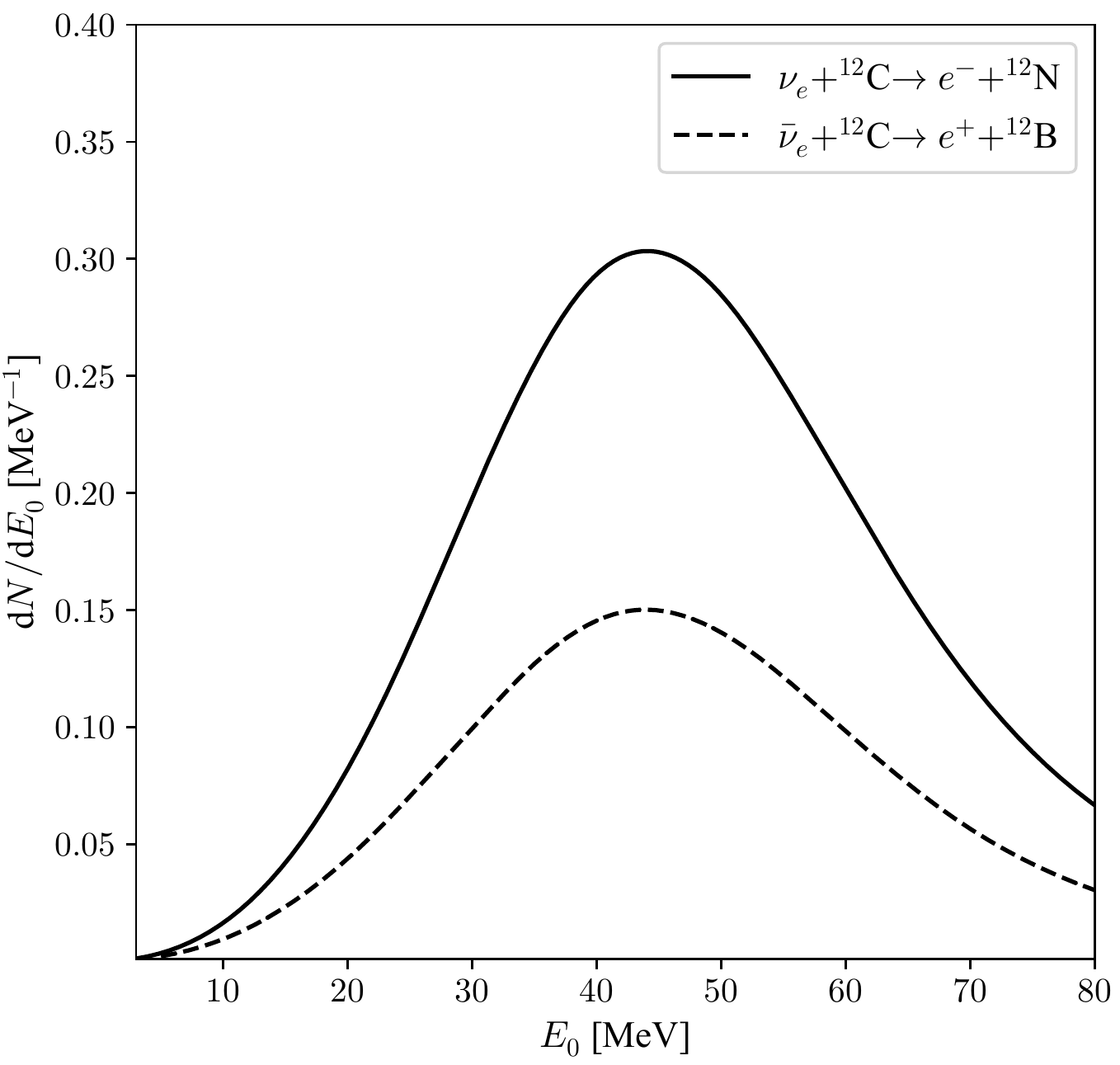}
    \caption{The event spectra of the \ac{CC} reactions with ${^{12}}{\rm C}$ for the \acp{PBH} of $M^{}_{\rm PBH} = 10^{15}~\text{g}$, $f^{}_{\rm PBH}=5.5\times 10^{-4}$, where the effective running time of $10$ years and the fiducial \ac{LS} mass of $20$ kton are taken.}\label{fig:1e15_C12_CC}
\end{figure}

Similar to the $p$ES channel, only high-energy neutrinos and antineutrinos can produce the signals with energies observable in the \ac{LS} detectors. Therefore, we consider only the \acp{PBH} with the lightest mass, i.e., $M^{}_\text{PBH} = 10^{15}~\text{g}$. The event spectra of \ac{NC} and \ac{CC} reactions are shown in \citefig{fig:1e15_C12_NC} and \citefig{fig:1e15_C12_CC}, respectively, where the effective operation time of $10$ years and the fiducial \ac{LS} mass of $20$ kton have been taken as before. In comparison with the IBD channel, the event rates in the $^{12}$C channel are much smaller and will not contribute much to the detection of \acp{PBH}.

\section{Constraints from JUNO}\label{sec:sensitivity}
In the previous discussions, we have demonstrated that the IBD channel is most sensitive to the antineutrino signals from the \acp{PBH} as dark matter. For this reason, only the IBD signal will be implemented to derive the constraints on the \acp{PBH} at JUNO in this section. By requiring the signal-to-noise ratio $\text{\ac{SNR}} \ge 1.6/\sqrt{N}$ (i.e. 90\% C.L.), where $N$ is the total event number of backgrounds for 10 years, we have shown the upper limit on $f^{}_\text{PBH}$ for a given $M^{}_\text{PBH}$ in \citefig{fig:contrast}. The solid curve stands for the constraint from \ac{JUNO} in the case of Majorana neutrinos. For comparison, the currently best limit on $f^{}_\text{PBH}$ from \ac{Super-K} has been plotted the dot-dashed curve and the shaded area in light grey has been excluded \cite{Dasgupta:2019cae}. Two important observations can be made. First, after running for 10 years, \ac{JUNO} will have the capability to explore the dark grey area in \citefig{fig:contrast}, which has not been constrained by current neutrino observatories. To be specific, given $M^{}_{\rm PBH} = 10^{15}~{\rm g}$, the upper limit $f^{}_{\rm PBH} \lesssim 3\times 10^{-5}$ can be obtained from \ac{JUNO}, which is about twenty times better than that $f^{}_{\rm PBH} \lesssim 6\times 10^{-4}$ from \ac{Super-K}. Second, \ac{JUNO} will be able to constrain the \ac{PBH} dark matter in the mass range $M^{}_{\rm PBH} = (5 \sim 8)\times 10^{15}~\text{g}$, for which $f^{}_{\rm PBH} = 1$ is still allowed by \ac{Super-K}. As we have mentioned before, although there exist other observational limits on $f^{}_{\rm PBH}$ from cosmic gamma rays, the neutrino observations will not only provide an independent limit on $f^{}_{\rm PBH}$ but also a novel way to probe the mechanism of Hawking radiation of \acp{PBH}.

\begin{figure}
    \includegraphics[width = 0.45\textwidth]{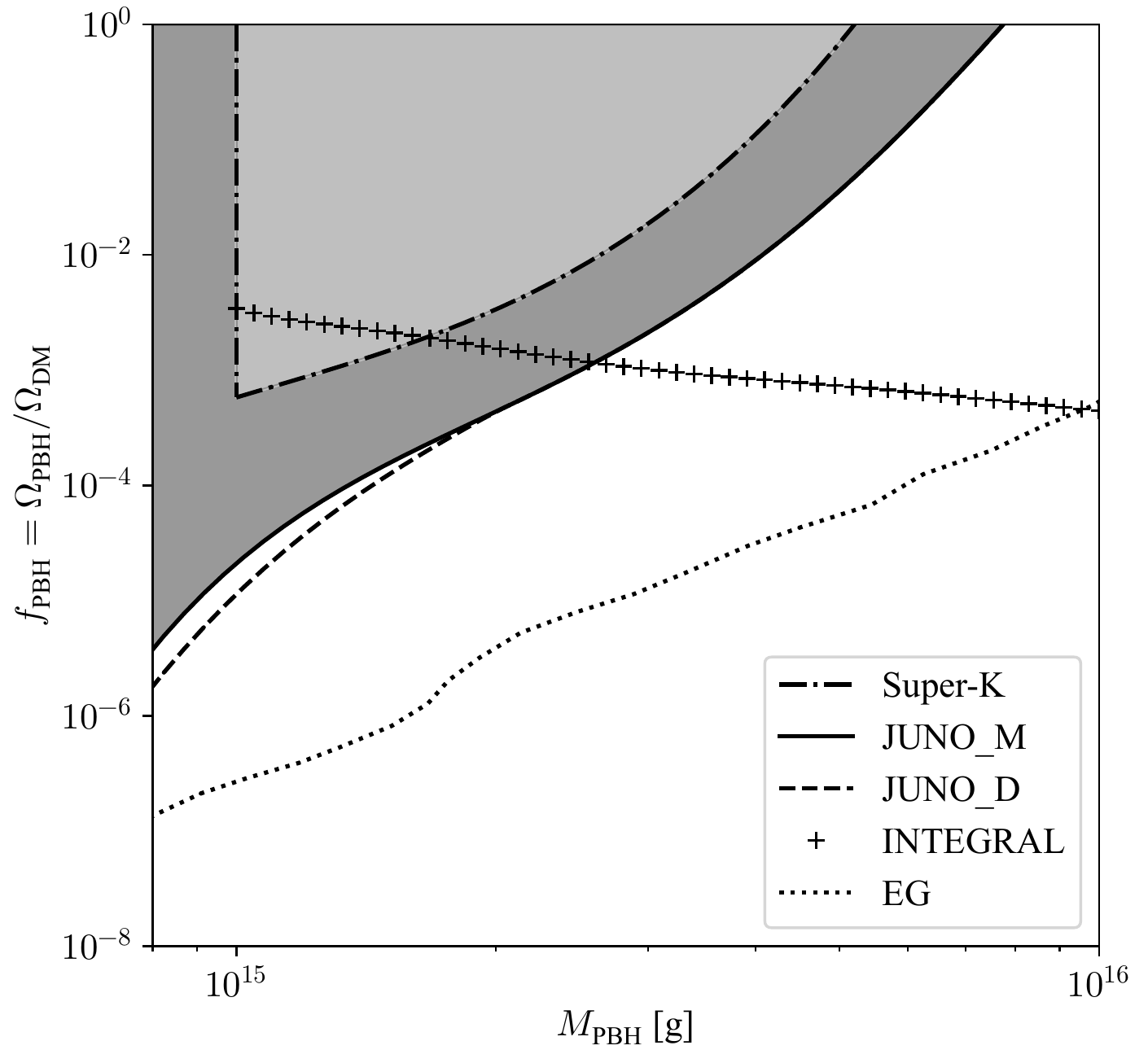}
    \caption{The expected upper limit on the \ac{PBH} abundance $f^{}_{\rm PBH} \equiv \Omega^{}_{\rm PBH}/\Omega^{}_{DM}$ for a given \ac{PBH} mass from JUNO with an effective running time of 10 years and a fiducial \ac{LS} mass of $20$ kton.
    The solid and dashed curves stand for the upper limits from JUNO in the case of Majorana and Dirac neutrinos, respectively. The dot-dashed curve denotes the upper limit from Super-K \cite{Dasgupta:2019cae}. 
    The cruciate-pixels curve is the constraint from the $511$keV $\gamma$-ray line due to $e^-$-$e^+$ annihilation measured by \ac{INTEGRAL} \cite{Dasgupta:2019cae}. The dotted curve denotes the constraint from the extragalactic $\gamma$-rays (EG) \cite{0912.5297}.}\label{fig:contrast}
\end{figure}

Now we consider the difference between Majorana and Dirac neutrinos. The key point is that the evaporation rate of PBHs depends on the total number of degrees of freedom in the neutrino sector, which is $n^{\rm M}_{\rm dof} = 6$ for Majorana neutrinos and $n^{\rm D}_{\rm dof} = 12$ for Dirac neutrinos. Notice that three generations of massive neutrinos and two helical states for each generation have been taken into account. Furthermore, neutrinos and antineutrinos have to be distinguished in the case of Dirac neutrinos. Although the right-handed neutrinos and left-handed antineutrinos in the Dirac case don't participate in ordinary weak interactions, they will be produced in the Hawking radiation. Therefore, similar to Majorana neutrinos, we have recalculated the neutrino fluxes and event rates for Dirac neutrinos, and plotted the projected constraint on the PBH adundance in Fig.~\ref{fig:contrast} as the dashed curve. Comparing between the constraints in the Majorana case (solid curve) and the Dirac case (dashed curve), we can make two interesting observations. First, for the PBH masses $M^{}_{\rm PBH} > 2\times 10^{15}~{\rm g}$, there is essentially no difference between the Majorana and Dirac cases. The main reason is that the evaporation rate is so small that the PBH mass is reduced by less than $3\%$ from the formation time to the present. Second, the constraint becomes more restrictive in the Dirac case for $M^{}_{\rm PBH} \lesssim 2\times 10^{15}~{\rm g}$. For instance, the relative difference between the constraints for Majorana and Dirac neutrinos could reach $40\%$ for $M^{}_{\rm PBH} = 10^{15}~{\rm g}$. This can be understood by noticing that the difference in the number of degrees of freedom turns out to be important when the evaporation rate becomes significantly large. As a consequence, the production rate of Dirac neutrinos is higher than that of Majorana neutrinos.

\section{\label{sec:con}Concluding Remarks}

Motivated by the attractive scenario of \ac{PBH} dark matter, we have calculated the \ac{PBH}-induced neutrino and antineutrino fluxes due to the Hawking radiation and the corresponding neutrino event spectra at \ac{JUNO}. The main conclusion is that the current limit on the \ac{PBH} abundance $f^{}_{\rm PBH}$ from \ac{Super-K} will be improved by one order of magnitude at \ac{JUNO}. 

One should notice that \acp{PBH} have been assumed to follow the monochromatic mass distribution and to be spinless. However, there are indeed different theoretical scenarios, in which the predicted mass distribution of \acp{PBH} is broad \cite{Carr:1975qj, Harada:2016mhb, Kannike:2017bxn, Yokoyama:1998xd} or the \ac{PBH} spins are high \cite{Harada:2017fjm, Arbey:2019jmj}. For the \acp{PBH} with a log-normal mass distribution, the shaded regions in \citefig{fig:contrast} would be shallower but broader (e.g., see Fig.~2 in Ref.~\cite{Dasgupta:2019cae}). For the spinning \acp{PBH}, the constraints would be more stringent because of more intense Hawking radiation \cite{Macgibbon1990Quark,Macgibbon1991Quark}. Therefore, the exclusion limits obtained in the present work can be regarded to be conservative. Our approach can be easily generalized to study more generic scenarios of \acp{PBH}, which is however beyond the scope of this paper.

As shown in \citeeq{eq:Gal}, a different choice of the density profile of dark halo may affect the neutrino fluxes from the Galaxy. Besides the \ac{NFW} profile, we have also considered two other typical scenarios, i.e., the \ac{ISO} profile and the \ac{EIN} profile \cite{ng_resolving_2014}. Define the relative error as 
\begin{equation}\label{eq:error}
    \mathrm{err}(\mathrm{X})\equiv\frac{|{\rm d}F^{}_{\mathrm{Gal}}/{{\rm d}E}(\mathrm{X})-{\rm d}F^{}_{\mathrm{Gal}}/{{\rm d}E}(\mathrm{NFW})|}{{\rm d}F^{}_{\mathrm{Gal}}/{{\rm d}E}(\mathrm{NFW})}\ .
\end{equation}
We find that its magnitude is about $1.6\%, 0.3\%$ for $\mathrm{X}=\mathrm{\ac{ISO}}$ and \ac{EIN}, respectively. These errors are small enough for us to safely ignore the uncertainty from different choices of the density profiles. Therefore, we just take the \ac{NFW} profile in the present work for illustration.

Neutrino flavor conversions may also affect the detection of neutrinos from PBHs at JUNO. Given the neutrino $\nu^{}_\alpha$ emitted from the PBHs, the probability to have $\nu^{}_\beta$ at the detector is given by \cite{Zhi2011Neutrinos}
\begin{equation}\label{eq:prob}
    \begin{split}
        P(\alpha\rightarrow\beta)=&\sum_{j=1}^3|U_{\alpha j}|^2 |U_{\beta j}|^2\\
        &+2\mathrm{Re}\sum_{j<k}U_{\alpha j}U_{\beta k}U_{\alpha k}^\ast U_{\beta j}^\ast e^{-i\frac{\Delta m_{kj}^2}{2E}L}\\
        \approx&\sum_{j=1}^3|U_{\alpha j}|^2 |U_{\beta j}|^2 \ ,
    \end{split}
\end{equation}
where $U^{}_{\alpha j}$ (for $\alpha = e, \mu, \tau$ and $j = 1, 2, 3$) denote the elements of the lepton flavor mixing matrix \cite{Maki:1962mu}, $\Delta m^2_{kj} \equiv m^2_k - m^2_j$ (for $kj = 21, 31, 32$) are neutrino mass-squared differences, $E$ is the neutrino energy, and $L$ is the distance between the neutrino source and the detector. In Eq.~(\ref{eq:prob}), the complete decoherence of the neutrino state has been assumed such that the oscillation terms disappear, as the distance $L$ between the source to the detector is much longer than the neutrino oscillation length $L^{}_{\rm osc} \equiv 4\pi E/\Delta m^2_{21} \approx 3.3 \times 10^{5}~{\rm m}$ for $E = 10~{\rm MeV}$ and $\Delta m^2_{21} = 7.5\times 10^{-5}~{\rm eV}^2$. Then, the neutrino flux at the detector is related to the original flux via
\begin{eqnarray}
\frac{{\rm d}F^{\rm d}_{\nu^{}_\alpha}}{{\rm d}E} = \sum_{\beta} \frac{{\rm d}F^{}_{\nu^{}_\beta}}{{\rm d}E} P(\nu^{}_\beta \to \nu^{}_\alpha) \; ,
\label{eq:fluxosc}
\end{eqnarray}
where the oscillation probability $P(\nu^{}_\beta \to \nu^{}_\alpha)$ has been given in Eq.~(\ref{eq:prob}) and it is also applicable to antineutrino oscillations. By taking account of neutrino oscillations, we have numerically checked that the corrections to the event spectra in the IBD and $e$ES channels are below $2.5\%$, which will not significantly affect the constraints shown in Fig.~\ref{fig:contrast}. In addition, the $p$ES is induced by neutral-current interactions, so it is insensitive to neutrino oscillations.

Another concern is related to the $p$ES channel for the detection of neutrinos and antineutrinos from \acp{PBH}. Compared to the IBD channel, a considerable event rate produced from \acp{PBH} is also expected in this channel. However, we have shown that it is severely contaminated by the backgrounds, which have not been well studied so far. The derived limit from JUNO in \citefig{fig:contrast} will be further improved if new analysis techniques are developed to effectively get rid of such backgrounds \cite{1507.05613}. This exploration will be left for future works. 

\section*{Acknowledgments}
The authors are greatly indebted to Liang-Jian Wen for helpful discussions. This work was supported in part by the National Natural Science Foundation of China under Grants No. 11675182 (Z.C.), No. 11690022 (Z.C.), No. 11775232 (S.Z.) and No. 11835013 (S.Z.); by the CAS Center for Excellence in Particle Physics (S.Z.); and by a grant for S.W. under Grant No. Y954040101 from the Institute of High Energy Physics, Chinese Academy of Sciences.

\bibliography{biblio}

\end{document}